\documentstyle[aps,prb,psfig,multicol]{revtex}

\renewcommand{\vec}[1]{{\bf #1}}

\newcommand{\nought}{{\rm o}}

\newcommand{\be}{\begin{equation}}
\newcommand{\ee}{\end{equation}}
\newcommand{\bea}{\begin{eqnarray}}
\newcommand{\eea}{\end{eqnarray}}

\newcommand{\ea}{\mbox{\em et al.}}

\def\ybcoAB{Y$_2$Ba$_4$Cu$_7$O$_{15}$}
\def\ybcoA{YBa$_2$Cu$_3$O$_{7}$} 
\def\ybcoB{YBa$_2$Cu$_4$O$_{8}$}

\newcommand{\rfig}[1]{Fig.~\ref{#1}}

\newcommand{\cut}[1]{}

\begin{document}
\draft
\title{%
Interplane magnetic coupling effects in 
the  multilattice compound \ybcoAB 
}
\author{G.~Hildebrand$^1$, E.~Arrigoni$^1$, J.\ Schmalian$^2$, and W. Hanke$^1$}
\address{$^1$Institut f\"ur Theoretische Physik, 
Universit\"at W\"urzburg, Am
Hubland, D-97074 W\"urzburg, Germany}

\address{%
$^2$Department of Physics, University of Illinois at
Urbana-Champaign, 1110 W. Green Str., Urbana 61801, IL}

\maketitle
\begin{abstract}

We investigate
the  interplane magnetic coupling of the
multilattice compound \ybcoAB
by means of a bilayer Hubbard model with inequivalent planes.
 We evaluate the spin response, effective interaction and
the intra- and interplane spin-spin relaxation times
 within the fluctuation exchange
approximation.
We show
that strong in-plane  antiferromagnetic fluctuations 
are responsible for a magnetic coupling
 between the planes, which in turns leads to a
tendency of the fluctuation in the two planes to equalize.
 This equalization effect grows 
whit increasing in-plane antiferromagnetic fluctuations, i. e., with 
 decreasing temperature and
decreasing doping, while it is completely absent when the in-layer
correlation length becomes of the order of one lattice spacing. 
Our results provide a good qualitative description
of NMR and NQR experiments in \ybcoAB.

\end{abstract}
\pacs{PACS numbers: 74.72.-h,71.27.+a,76.60.-k
\hfill{\bf to appear in Phys. Rev. B (RC) Jan. 99}
}

\ifpreprintsty\onecolumn\fi


\begin{multicols}{2}

Although many models for high-Tc cuprates are restricted to a
 single layer, 
it has become clear that both superconducting and magnetic properties
of these materials
 are  affected by the coupling between two or more layers.
A  rather strong  coupling between 
the layers  has been observed  principally by inelastic neutron
scattering~\cite{tr.ge.92}   (INS) and nuclear magnetic 
resonance~\cite{st.ma.94,st.ma.95.1,st.ma.95.2,mo.ri.94} (NMR).
Furthermore, the observation of a qualitatively different behavior
of the odd and even channel in INS~\cite{bo.fo.97} 
and  of a bilayer 
splitting of the Fermi surface  found in  angular resolved photoemission 
experiments (ARPES)~\cite{sc.pa.98.1,sc.pa.98.2}
demonstrate that   low energy excitations 
of   cuprates are affected by the presence of more than one
layer per unit cell.
An exciting   perspective on the nature of the coupling between 
CuO$_2$-layers  was offered by NMR experiments  by Stern {\em et al.}
on \ybcoAB\ (247). This  material  has a variety of
structural similarities to the extensively studied  
 \ybcoA\ (123) and \ybcoB\ (124) systems.
The compound 247 can be considered as a natural multilattice, 
whose bilayers  are build up of one
CuO$_2$ layer which belongs to the 123 block and one
 layer to the 124 block.
Based on the analysis of the  NQR  spectra it turned out that the 
charge carrier content in these nonequivalent adjacent layers
is very close to that of the related parent compounds of the two
 blocks, 123 and 247.
Interestingly, the highest transition temperature (T$_c = 95 \, {\rm K}$)
occurs in the
 247 compound, in comparison with the $92 \, {\rm K}$
of 123 and $82 \, {\rm K}$ of the 124 system.

In this paper, we want to provide a theoretical understanding in terms
of a microscopic model of some 
striking experimental  observations of
Refs.~\onlinecite{st.ma.94,st.ma.95.1}, namely:
(i)
the spin-spin relaxation rates T$_{2G}^{-1}$ of the two layers in \ybcoAB,
measured in a spin-echo
double resonance experiment,
 behave very similarly as a function of temperature,
 despite the
different doping of the layers; (ii) 
the 
spin-spin relaxation rate in the 124 (247) layer of \ybcoAB
is reduced (enhanced)
with respect to one of the
constituent compound at low temperatures; 
(iii)
the interplane transverse relaxation rate, 
 increases for decreasing temperature faster
than the intraplane one.
The overall features that can be inferred from these experiments 
are that,
for high temperatures
the magnetic fluctuations in the two layers of \ybcoAB are  
disconnected and each layer behaves similarly 
to the corresponding parent compounds,
whereas for  decreasing temperatures, the increasing interlayer
magnetic coupling enforces
even the slightly overdoped plane to behave like an underdoped system
and vice-versa.

To describe the strong electronic correlations in the system \ybcoAB,
consisting of two layers with different charge carrier concentration,
we extend the standard single-band Hubbard model to include two layers 
coupled by an interplane hopping matrix element $t_\perp$.
Furthermore, to produce a different hole concentration in the
planes, we introduce an on-site energy $\delta$ in the second plane.
After Fourier transformation of the kinetic part 
the Hamiltonian  reads:
\be
H= \sum_{\stackrel{l_1,l_2}{\vec{k},\sigma}} \left[ H_\nought(\vec k)
\right]_{l_1,l_2} \cdot
c^\dagger_{\vec{k}, l_1,\sigma}
c^{}_{\vec{k}, l_2,\sigma}
+ U \sum_i n_{i, \uparrow}  n_{i, \downarrow} \; .
\label{eq:Hubbard_Hamiltonian}
\ee
where 
$c^\dagger_{\vec{k}, l_1,\sigma}$ creates a particle with spin $\sigma$ and
momentum $\vec k$ in layer $l_1$, and $i$ runs over the sites of the
two layers.
In contrast to the standard notation in the
monolayer case,  $\epsilon_{\vec k}$ is replaced by the $2\times 2$
matrix.
 $H_\nought(\vec k)$. 
whose components are 
$[H_\nought(\vec k)]_{11}=\epsilon_{\vec k} -\mu$, $[H_\nought(\vec
k)]_{22}=\epsilon_{\vec k} +\delta -\mu$ and $[H_\nought(\vec
k)]_{12}=[H_\nought(\vec k)]_{21}=t_\perp$.
The bare dispersion $\epsilon_{\vec k}$ includes second $(t')$ and third
$(t'')$ nearest-neighbor hopping processes to better model the Fermi
surface of the cuprates (see, e.g. \cite{to.ma.94,ki.wh.98}). 
The properties of the interacting system are deduced from the Green's
function and form the dynamic two-particle susceptibilities 
in the framework of the fluctuation exchange approximation
(FLEX) \cite{bi.sc.89}. 
In this approximation, bubble and ladder diagrams are summed up in
infinite order and the resulting coupled set of equations is solved
self-consistently.
Although this approximation is suitable to describe magnetic
 fluctuations and band shadow features,\cite{la.sc.95} it fails to
 reproduce the pseudogap behavior 
in the spectral function and in the magnetic
 excitations at low temperatures and low doping\cite{vilk},
probably due to the
lack of conservation at the two-particle level and to the
neglection of interference effects between  particle-particle and
particle-hole channel.
Moreover, on the quantitative level, antiferromagnetic fluctuations
appear to be 
underestimated\cite{hi.ar.98.m1} with respect to exact results.

In order to understand the systematics and the parameter dependence of
the interplane magnetic coupling,
we first focus our attention on two systems with different 
parameter sets $t'/t,t''/t$
corresponding to  strong and weak antiferromagnetic fluctuations.
Specifically, we introduce a first system, for simplicity labeled by
``A'', with  
$t'=-0.38t, t''=-0.06t$, , and a second one (``B'') with $t'=-0.20t,
t'=0.15t$. System A
shows much stronger antiferromagnetic fluctuations
due to the underlying Fermi surface  in comparison to B.
 The Hubbard interaction  takes an intermediate value 
$U=4t$, as appropriate for a perturbative calculation, 
and $t_\perp=0.4t$ (cf. \cite{gr.sc.97.1,gr.sc.97.2,hi.ar.98.m1}).
Within
 the self-consistency cycle, we fix  the on-site energy $\delta$
and the particle number $n_1 = 1 - x_1$ of the first layer, while the
chemical potential $\mu$ and the particle number of the second plane
$n_2=1-x_2$ are adjusted at each step.
Keeping this way  the doping of the first plane fixed and changing  only
 the doping of the second one, we investigate a possible magnetic 
 coupling  between the layers.
Our question is whether the
 magnetic fluctuations of the first plane
are influenced by the doping of the second plane and vice-versa.

\begin{figure}
\centerline{\psfig{file=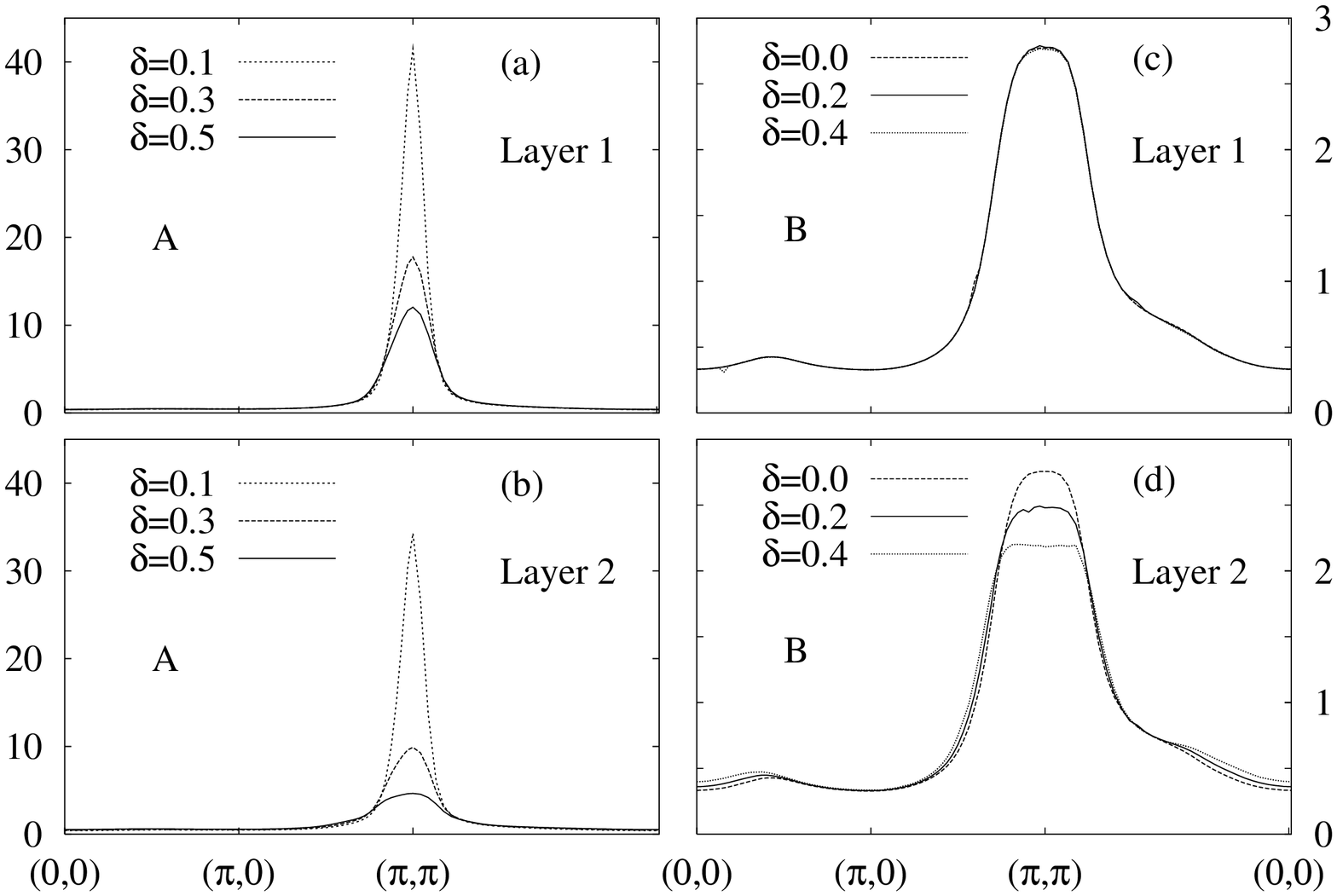,width=9cm}}
\caption{\label{fig:ReXband}%
\narrowtext
Static spin susceptibility 
$\chi^{zz}_{ll}(\vec q,\omega=0)$
along the standard path in the Brillouin zone
for the two layers and both systems A and B (see text). (The
temperature is $T=0.02t$, and $x_1=0.08$)}
\end{figure}

With this in mind,
 we consider in \rfig{fig:ReXband}
the static spin-spin function $\chi^{zz}_{ll}(\vec q,
\omega=0)$ along the standard path $(0,0) \to (\pi,0) \to (\pi,\pi) \to
(0,0)$ in the Brillouin zone for the two layers ($l=1$ and $l=2$) and both
systems A and B.
\vspace*{-.3cm}
As one can clearly see
from Fig.~\ref{fig:ReXband}(a) and (b)
 for the case of system A,
the spin response in the two planes is strongly peaked at the
antiferromagnetic wave vector $\vec Q=(\pi,\pi)$ indicating pronounced 
antiferromagnetic fluctuations in the Hubbard planes.  
 The  striking result is
that data for a plane with a given doping ($l_1=1$) also depend on whether
this plane is coupled with an equivalent one  or with a
more or less doped one, depending of the value of $\delta$.
 This behavior clearly suggests a strong magnetic 
connection between the planes.
The decrease of the spin susceptibility with increasing $\delta$
[see \rfig{fig:ReXband}(a) and (b)] is due to the increased hole
concentration of the total system moving it further away from
half filling where antiferromagnetism is most pronounced.
The situation is rather different for system B, whose data are
shown in 
\rfig{fig:ReXband}(c) and (d).
Here, a variation of $\delta$ influences  the susceptibility of the 
second plane only, whereas the 
first plane, with constant charge carrier concentration, 
 is almost not affected at all.  Thus, the magnetic fluctuations in the 
first plane are   
disconnected from the fluctuations in the second one.
Only for a system with sufficiently strong antiferromagnetic correlations
and a correlation length larger than a few lattice constants, the two inequivalent layers turn out
to be strongly coupled, as in case A.
\begin{figure}
\vspace*{-.5cm}
\psfig{file=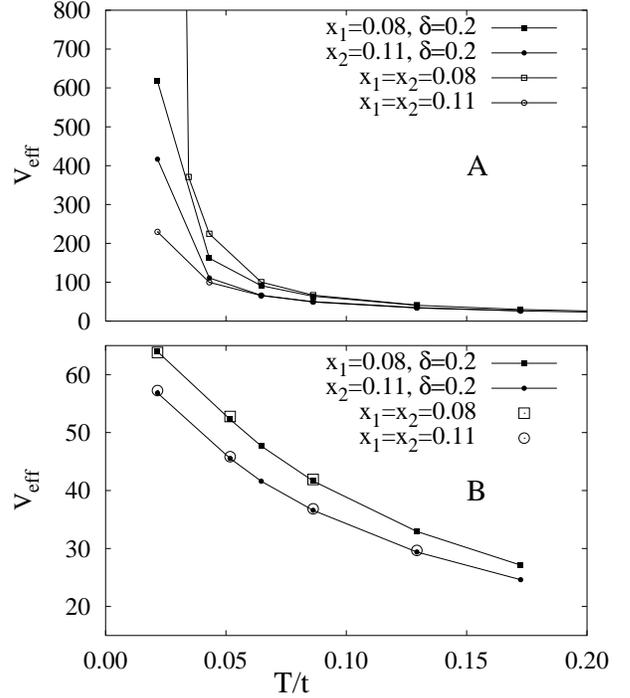,width=8cm}
\vspace*{-.2cm}
\caption{\label{fig:Veff}%
Effective interaction $V_{\rm eff}(\vec Q,\omega=0)$ vs. $T$
for a  system  with 
inequivalent layers ($x_1=0.08$ and $x_2=0.11$) in comparison
with the corresponding
bilayer systems with $x_1=x_2=0.08$ and $x_1=x_2=0.11$.
The upper panel shows the results
for system A and the lower for  B.}
\end{figure}
\vspace*{-.3cm}

We now ask  to what extent each layer of a 
  system with inequivalent layers
compares to the corresponding system with equivalent layers (i.e.,
$\delta=0$). Therefore, we show in \rfig{fig:Veff}
the effective interaction $V_{\rm eff}(\vec Q,\omega=0)$ 
at the antiferromagnetic wave vector $\vec Q=(\pi,\pi)$ as a function of
temperature for systems with inequivalent layers $(x_1=0.08, x_2=0.11,
\delta=0.2)$ in comparison with the two corresponding systems with
identical layers $(x_1=x_2=0.08)$ and $(x_1=x_2=0.11)$.
In the upper panel of \rfig{fig:Veff} we show the results for 
 system A.
Here, 
we clearly observe  considerable differences between each of 
the layers  in the inequivalent-layer system with respect 
to its counterparts in the
equivalent-layer system at low temperatures.
Specifically, the effective interactions in the two layers
of the system with inequivalent layers are ``enclosed'' between the
ones in the corresponding layers of the system with identical layers. 
Thus, the interlayer magnetic correlations lead to
a tendency for the magnetic fluctuations of the two 
planes to equalize although their doping is different.
The results for  system B  are strikingly different
as can be seen  in the lower panel of
\rfig{fig:Veff}.
 Here, the two planes with
different dopings in the inequivalent-layer system behave essentially like
the correspondingly doped layers in the system with equivalent layers. 
For this less magnetic system the  fluctuations
 in the two layers are
effectively decoupled and do not influence  each other.
The results displayed in \rfig{fig:Veff} thus 
again   show that the antiferromagnetic
correlations within the layers 
 lead to a partial equalization of the 
magnetic fluctuations in the two planes.

A similar magnetic equalization  effect 
has  been observed  
by Scalettar et al. \cite{sc.ca.94}
in Quantum-Monte-Carlo
simulations of an half-filled layer coupled to a doped one.
According to these authors,  this equalization effect
can be explained
by perturbation in $t_\perp$ as a competition between 
 virtual hopping processes between the two layers with energy scale
 $\propto t_\perp/\delta$ and exchange energy $J\propto t^2/U$.
We believe that our use of the FLEX approximation
is essential in order to 
additionally explain the
crossover from the ``equalized'' regime (A) to the disconnected one (B),
which accounts for
the qualitative behavior of the
experiments on \ybcoAB\cite{st.ma.95.1} as we will show below.
This is due to
the feedback effect of
the Green's functions renormalization in the FLEX approximation.

The equalization effect observed in system A qualitatively describes
the experimental situation for the  inequivalent-layer system \ybcoAB.
Indeed,  NMR experiments by
 Stern et al. \cite{st.ma.95.1}
measuring $1/T_{2G}$ for the two layers of \ybcoAB\ and for the two corresponding
systems with equivalent layers, \ybcoA\ and
\ybcoB, show a similar equalization tendency as  in Fig.~
\ref{fig:Veff}.
To further demonstrate the similarities of our calculations with the
experimental results, we also evaluate  the nuclear
spin-spin relaxation time $T_{2G}$ for spins within the two planes.
The  generalized relaxation rate $T_{2G}^{ll'\ -1}$ measuring the
interaction between a spin in plane $l$ 
and one in $l'$ is related to a weighted
sum over the static spin susceptibility $\chi^{zz}_{ll}(\vec
q,\omega=0)$ where the momenta $\vec q \cong \vec Q$ have the strongest
weight.\cite{mo.ri.94,th.pi.94,taki.94,mi.mo.96}.
In our calculations, we assume that the hyperfine coupling constants
entering the form factor are the same for both planes and use the values
given by Barzykin and Pines \cite{ba.pi.95}. Hence possible difference
in the relaxation times of the two planes can  solely arise from different spin
responses in the layers. 
In a previous work on \ybcoAB, Millis and Monien,
who  determined the value  of the interlayer exchange interaction from the 
experimental data of Ref.~\cite{st.ma.95.1},
 have considered an alternative   point of view  and
neglected the difference between the 
spin
susceptibilities of the two layers by considering 
different hyperfine coupling constants
\cite{mi.mo.96}. This approach is 
justified
by our results which show for systems
with strong antiferromagnetic correlations a pronounced coupling of the magnetic response
of the two layers and thus a tendency to equalization of the
susceptibilities.
 This primarily causes a similar temperature dependence
of the two in plane susceptibilities. Nevertheless, 
quantitatively,  the   magnitudes
of the magnetic  response of the two layers  stays different.

\begin{figure}
\vspace*{-.5cm}
\psfig{file=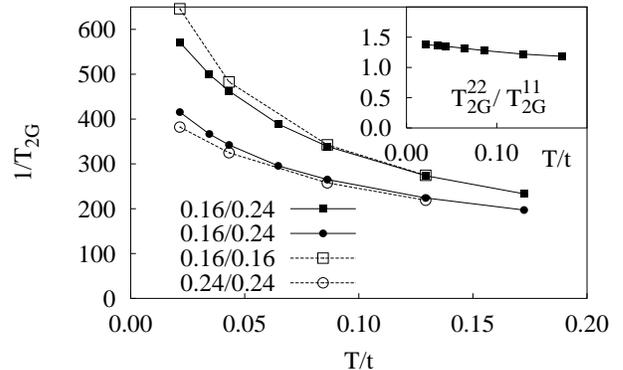,width=8cm}
\vspace*{-.2cm}
\caption{\label{fig:tdl_T2G}
Temperature dependence of the
Gaussian component of the spin-spin relaxation
rate $T_{2G}$ of a bilayer system consisting of two different layers
with doping
 $x_1=0.16$ and $x_2=0.24$, respectively
$(\delta=0.4)$.
For comparison, we show also the result
for two bilayer systems with equivalent layers with doping
$x_1=x_2=0.16$ and $x_1=x_2=0.24$, respectively. The other parameters
for all curves are:
$t'=-0.38t$,$t''=-0.06t$. The inset shows the ratio
$T^{22}_{2G}/T^{11}_{2G}$ for the case of inequivalent layers.}
\end{figure}
\vspace*{-.3cm}
In \rfig{fig:tdl_T2G} we show the in-plane
$1/T_{2G}$ as a function of temperature
for both planes of the system with {\it inequivalent} layers again
in comparison with the  data for two corresponding
bilayer systems with {\it equivalent} layers.
In this figure, the filled (open) symbols  represent the system with
inequivalent (equivalent) layers. Furthermore, the squares are related to the
data for planes with a  doping  $x=0.16$ and the bullets
planes with doping  $x=0.24$.
Since the main contribution to $1/T_{2G}$ stems from $\chi^{zz}(\vec
Q,0)$, a system with rather strong magnetism 
exhibits large relaxation rates $1/T_{2G}$. This explains why the
 layers with $x=0.16$  show a smaller relaxation rate compared to
$x=0.24$.
As expected form the results of \rfig{fig:Veff}, and as observed in
NMR experiments at lower temperatures\cite{st.ma.95.1}, the heavily 
doped plane $(x_2=0.24)$ of the 
system with inequivalent layers
shows stronger magnetic fluctuations than the plane in the corresponding
system with equivalent layers $(x_1=x_2=0.24)$. Similarly, the
magnetism of the lower doped plane $(x_1=0.16)$ is reduced
with respect to the corresponding equivalent-layer system $(x_1=x_2=0.16)$
due to the coupling to a more doped plane.
As already observed in the results for the effective interaction shown
above, 
the magnetic fluctuations of the
two inequivalent planes with different carrier concentration
tend to be partially equalized by interplane coupling effects.
The theoretical  results in \rfig{fig:tdl_T2G}, as well as the
experimental results, show a strong
increase of $1/T_{2G}$ with decreasing temperature. However, the
experimentally observed  
decrease of $1/T_{2G}$  below $T_{sg} \approx 100{\rm K}$
due to the 
 opening of a pseudogap in the spin excitation spectrum,
is not reproduced by the FLEX approximation, as discussed above.

As NMR experiments show,
 the spin-lattice relaxation rate $1/T_{2G}$  
has the same temperature dependence
in the two planes of \ybcoAB\cite{st.ma.95.1}, the
ratio
$R=(1/T^{124}_{2G})/(1/T^{123}_{2G})$ 
being temperature independent and 
approximately $R\approx$~1.4--1.5.
This ratio corresponds to $R=T^{22}_{2G}/T^{11}_{2G}$
in our work, 
since the 124 layer in the coupled layer structure
of \ybcoAB\ is the one with lower doping (here  labeled by ``1'') and 
the $\rm CuO_2$-layer from the 123 block corresponds to
the second plane in our theoretical study.
The theoretical values for $R$ are  shown in the inset of
 \rfig{fig:tdl_T2G} as a function of temperature.
For the parameter set chosen (corresponding to system A),
$R$ is approximately  $1.2-1.4$
and  almost temperature--independent, in
good agreement with the experimental finding.

\begin{figure}
\vspace*{-.5cm}
\psfig{file=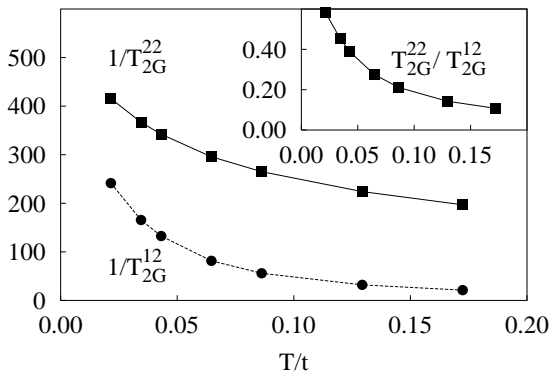,width=8cm}
\vspace*{-.2cm}
\caption{\label{fig:SEDOR_T2G}
Intra- and inter-plane spin-spin relaxation rates
$1/T_{2G}^{22}$ and $1/T_{2G}^{12}$, respectively, vs. $T$.
The inset shows the ratio between the two
relaxation rates $(x_1=0.16, x_2=0.24, \delta=0.4)$.
}
\end{figure}
\vspace*{-.3cm}
In order to have a measure of the interplane spin coupling in \ybcoAB, Stern
\ea\cite{st.ma.95.2}
have carried out 
 NQR spin-echo
double-resonance (SEDOR) measurements, 
as suggested by Monien and Rice\cite{mo.ri.94}.
 These  experiments allow the determination of 
the inter-layer spin-spin relaxation time $T_{2G}^{12}$.
The apparent feature in their results is that the inter-plane 
relaxation rate, although smaller than the in-plane one,  
increases faster for decreasing temperature,
as seen from the temperature dependence of
the ratio $R_{\rm SEDOR}(T)=T_{2G}^{22}/T_{2G}^{12}$.
For the sake of comparison, we present 
our theoretical results for the 
in-plane relaxation rate $1/T_{2G}^{22}$ together with the inter-plane
one $1/T_{2G}^{12}$ in \rfig{fig:SEDOR_T2G},
the parameters being the same 
as in \rfig{fig:tdl_T2G}.
The fact that $1/T_{2G}^{12}$ grows faster with decreasing temperature is even
more apparent in 
 the inset. Here,  we show  the
temperature dependence of the ratio between the two relaxation rates.
Our theoretical results thus
 reproduce the qualitative behavior observed
experimentally. However, for $T$ smaller than the spin gap 
the experiments show a saturation effect, which obviously cannot be 
reproduced within
our approximation.

In summary, we have studied the magnetic interplane coupling 
of  the multilattice compound \ybcoAB by modeling it with
 two inequivalent Hubbard layers
 coupled by an interlayer hopping
$t_\perp$.
If the antiferromagnetic correlation length is less 
than 1--2 lattice spacings,
which happens for high temperatures or for a Fermi surface with
suppressed nesting,
we find that the two inequivalent layers are  disconnected
and keep their individual magnetic properties.
 However, once the antiferromagnetic correlations in the layer
with smaller charge carrier concentration is sufficiently large, the 
single-particle excitations
for momenta close to the Fermi surface and, in particular,
around the  momenta close to $(\pi,0)$ 
 as well as the magnetic excitations of
the two layers are strongly connected.
In this situation, the whole system reacts
magnetically
as a single system,
 despite its inhomogeneous charge density. 
Due to the strong magnetic correlations in the underdoped  layer the
magnetic in-plane order in the nominally overdoped layer is stabilized
via interplane magnetic coupling.
Thus, the layer with  lower charge carrier concentration acts 
like
an  external staggered field. 
For suitable Fermi surfaces and low enough
temperatures,
this mechanism makes 
the magnetic dynamics of both planes indistinguishable.
By choosing parameters which lead to a moderate equalization effect, we have
qualitatively reproduced the salient features of the in-plane and
out-of-plane 
spin-spin relaxation times observed in NMR measurements on \ybcoAB, as
compared with the ones of its constituent compounds.

This work has been supported in part by the Science and Technology
Center for Superconductivity through NSF-grant DMR91-20000, 
   the Deutsche Forschungsgemeinschaft (J.S.), the 
EC-TMR
  program  ERBFMBICT950048 (E.A.), and by FORSUPRA (G.H. and W.H.).
It is our pleasure to thank   D. Pines, R. Stern, 
H. Monien,
 M. G. Zacher, and
R. Eder
for  helpful discussions. The calculations were performed at the ZAM in
J\"ulich and  the LRZ in Munich.

%
\vspace*{-.2cm}
\bibliography{myrefs,preprints,mypublications,localbiblio,refs}  
\bibliographystyle{myprsty} 
\end{multicols}

\end{document}